\documentclass{article}

\usepackage{spconf}
\usepackage{amssymb}
\usepackage[ruled,vlined,linesnumbered]{algorithm2e}
\usepackage{amsthm}
\usepackage{amsfonts}
\usepackage{amsmath}
\usepackage{booktabs}
\usepackage{bm}
\usepackage{color}
\usepackage{cite}
\usepackage{comment}
\usepackage{epsfig}
\usepackage{euscript}
\usepackage{fancybox}
\usepackage{graphicx}
\usepackage{hyperref}
\usepackage{lipsum}
\usepackage{pgf}
\usepackage{psfrag}
\usepackage{pifont}
\usepackage{mathrsfs}
\usepackage{multirow}
\usepackage{setspace}
\usepackage{stfloats}
\usepackage{tabularx}
\usepackage{textcomp}

\title{Option Pricing via Multi-path Autoregressive Monte Carlo Approach}
\name{Wei-Cheng Chen and Wei-Ho Chung}
\address{Research Center for Information Technology Innovation\\
	Academia Sinica, Taipei, Taiwan\\
	Email: jimmyweicc@iis.sinica.edu.tw, whc@citi.sinica.edu.tw}

\begin{document}
\ninept
\maketitle

\begin{abstract}
The pricing of financial derivatives, which requires massive calculations and close-to-real-time operations under many trading and arbitrage scenarios, were largely infeasible in the past.
However, with the advancement of modern computing, the efficiency has substantially improved.
In this work, we propose and design a multi-path option pricing approach via autoregression (AR) process and Monte Carlo Simulations (MCS).
Our approach learns and incorporates the price characteristics into AR process, and re-generates the price paths for options.
We apply our approach to price weekly options underlying Taiwan Stock Exchange Capitalization Weighted Stock Index (TAIEX) and compare the results with prior practiced models, e.g., Black-Scholes-Merton and Binomial Tree.
The results show that our approach is comparable with prior practiced models.\footnote{This work was supported in part by the Ministry of Science and
Technology, Taiwan under grant numbers 104-2221-E-001 -008 -MY3,
105-2221-E-001 -009 -MY3, and
106-2218-E-002 -014 -MY4.} 

\end{abstract}

\section{Introduction} \label{sec:intro}
Derivative is a financial product whose value is determined by an underlying asset, such as stock, currency or commodity.
As a primary type of derivative, option is widely traded in financial market for a number of purposes, including speculation, hedging, spreading and creating synthetic positions.
According to investors' personal perspective for future economy, two types of option are traded on the market, i.e., Call Options and Put Options.
A call option is an agreement that gives an investor the right to buy the underlying asset at a specified price within a specific time period.
On the other hand, a put option is the right to sell a specified amount of an underlying asset at a specified price within a period of time.
Furthermore, option investors can enter into either long position or short position on both types of options under different trading strategies.
However, the major problem behind these trading purposes for option investors is to find option premium by using mathematical models.

The field of option pricing was first introduced during nineteenth centuries.
Related techniques used in this area can be classified into two aspects, namely closed-form solutions and numerical solutions.
In 1973, Black, Scholes \cite{BSM} and Merton \cite{1973-Merton-option} provided the state-of-art closed-form solution to European-style options pricing problem in their seminal studies, known as the Black-Scholes-Merton (BSM) model.
Their contributions have subsequently led to a thriving growth in options trading via a mathematical tractability and legitimacy for the traders and institutional regulators.
For example, Johnson and Shanno \cite{1987-Johnson-variancechanging}, and Hull and White \cite{1987-Hull-Stochastic} applied Stochastic Volatility model to do option pricing.
This model assumes that volatility has random processes and fluctuations over time.
Consequently, similar studies include those proposed by E. Stein and J. Stein \cite{1991-Stein-Stockpricestochastic}, and Heston \cite{1993-Heston-Stochasticvolatility}.
Moreover, Cox \cite{1975-Cox-CEV} derived the well-known Constant Elasticity of Variance model which is extended by Schroder \cite{1989-Schroder-CEVextended}.
However, closed-form solutions serve as a suitable model if financial instruments have simple structures and assumptions.
On the other hand, numerical methods such as Lattice approach, Monte Carlo Simulations (MCS) method, and Finite Difference method were commonly used to price derivatives with complex structure and products with path-dependent characteristic, e.g., Boyle \cite{1988-Boyle-Lattice}, Hull and White \cite{1988-Hull-lattice}, and Brennan and Schwartz \cite{1978-Brennan-DifferenceMethod}.
Since Boyle demonstrated how to price European-style options using MCS method\cite{1977-BOYLE-montecarlo}, MCS method has become one of the most important techniques in numerical solutions.
For instance, Boyle, Broadie and Glasserman \cite{1997-Boyle-Montecarlo-american} proved the ability to apply MCS method to evaluate American-style options.
Furthermore, Longstaff and Schwartz provided Least Squares Monte Carlo method to deal with option pricing when underlying asset follows a jump-diffusion process \cite{2001-Longstaff-Leastsquare}.
In the past, the major problem of numerical solutions was that the pricing results were not exact due to computational inefficiency; however, with the advancement of modern computing and related researches, e.g., Kim and Byun \cite{2015-Kim-parallelmontecarlo}, and Wang and Kao \cite{2016-Wang-parametermontecarlo}, this problem has substantially improved.

Recently, the growing trend of High Frequency Trading has led investors into seeking profit within a shorter period.
In consequence, short-duration options have substantially draw great attention because of their high volatility that provides investors a better opportunity to earn extra profit.
However, there is a paucity of literature on the discussion of short-term options pricing, e.g., Andersen, Fusari and Todorov \cite{2016-weekly}.
Hence, we would like to investigate the problem of options pricing when option period is less than or equal to a week.
In order to analyze the problem, we choose weekly options (TXOW) which issued by Taiwan Futures Exchange (TFE) and connected with Taiwan Stock Exchange Capitalization Weighted Stock Index (TAIEX).
Moreover, TXOW is an European-style option that can only be exercised at the option's expiration date.
In our approach, we utilize the correlation of day-by-day price change by designing a multi-path simulation algorithm to extract the information via autoregression (AR) process and MCS method \cite{1976-Box-Timeseries}.
Then, we compare our results with two kinds of commonly used models: (1) Black-Scholes-Merton (BSM) and, (2) Binomial Tree (BT).
In conclusion, the model we built (Multi-path Autoregression Monte Carlo Approach, MAMC) shows comparable performance as well as other commonly practiced models.

The rest of this paper is structured as follows: Section 2 describes the datasets used in this paper; Section 3 provides an overview of MAMC model; Section 4 contains description of both BSM model and BT model, and the indicators for performance measurement; Finally, the simulation results obtained in this work are examined in Section 5 and the conclusions of this paper are written in Section 6.

\section{DATA STRUCTURE} \label{sec:data}
In this paper, we use two sets of data that contain information of options and underlying assets.
First, we obtained a list of TXOW that was issued between January 7, 2015 to December 21, 2016.
Furthermore, ten call options and ten put options with strike price relatively close to prior trading day's TAIEX closing price, were selected on each issue date.
These information provide issue date, due date, strike price, and its market price on each trading day for each option.
Second, we collect a series of TAIEX's daily closing price from 2014 to 2016 through Google Finance API.
Details of chosen options are described in Panel A (2015) and Panel B (2016) of TABLE 1.

\begin{table}[h]\small
	\centering
	\caption{Details of option data}
		\label{tb:gaussian_model_test_result}
		\tabcolsep= 7.0pt
		\begin{tabular}{lcccccc}
			\toprule[2pt] &
			\multicolumn{3}{c}{Type} &
			\multicolumn{3}{c}{Moneyness}\\
			\cmidrule(r){2-4}
			\cmidrule(l){5-7}
			Panel A: 2015 & All & Call & Put & ITM & NTM & OTM\\
			\midrule
			Total & 794 & 397 & 397 & 317 & 160 & 317\\
			Percent (\%) & 100 & 50.0 & 50.0 & 39.9 & 20.2 & 39.9\\
			\bottomrule[1pt]

			\toprule[1pt] &
			\multicolumn{3}{c}{Type} &
			\multicolumn{3}{c}{Moneyness}\\
			\cmidrule(r){2-4}
			\cmidrule(l){5-7}
			Panel B: 2016 & All & Call & Put & ITM & NTM & OTM\\
			\midrule
			Total & 798 & 399 & 399 & 319 & 160 & 319\\
			Percent (\%) & 100 & 50.0 & 50.0 & 40.0 & 20.0 & 40.0\\
			\bottomrule[2pt]
		\end{tabular}\par\smallskip
		\parbox{8.8cm}{
		*  In-the-money (ITM) call/put options are options with strike price lower/larger than underlying asset's current price.\\
		** Near-the-money (NTM) call/put options are options with nearest strike price to underlying asset's current price.\\
		*** Out-of-the-money (OTM) call/put options are options with strike price larger/lower than underlying asset's current price.}		
\end{table}
\section{Multi-path autoregression algorithm}
\label{sec:algorithm}
\subsection{Model Overview}
In this section, we give the details of the construction and processing steps in our model.
First, we consider TAIEX as a discrete-time asset with closing price $S_t$ at time $t$, and $S_{t-1}$ at time $t-1$.
In order to estimate the regularities and patterns, we have to generate serial price return data with the rate of price changes. 
Therefore, we define the change rate $y_t$ as the logarithm of the ratio of closing price:
\begin{equation}
y_t = \ln\frac{S_t}{S_{t-1}},~~~t = 1, 2, 3,... .
\end{equation}
Hence, closing price at time $t$ can be represented as $S_t$.
That is:
\begin{equation}
S_t = S_{t-1} * \exp(y_{t}),~~~t = 1, 2, 3,... .
\end{equation}

Second, we assume that price return on each day consists of two parts: an expected return as a dependent variable affected by previous price change which generated through AR process and an unexpected return whose value are computed by performing MCS method on a stochastic process.
On top of that, we estimate both parameters by using underlying asset's closing price based on $N$-day rolling horizon statistics.
After describing the concept in our model, we apply this model to price the premium of TXOW which issues at time $t$.

\subsection{Expected Return}
Before pricing the option, we set a training dataset of price return $N$ days before time $t$ for regression process.
Then, we calculate logarithmic price changes with (1) and define price return series $Y_{N,t}$ as:

\begin{equation}
Y_{N,t} = \{ y_{t-i} | ~i = 1, 2, 3,..., N\}.
\end{equation}

After that, we compute first-order autocorrelation parameter $\alpha$ for entire $Y_{N,t}$ series under Least Square Method (LSM).
By using the computation results, we can write the expected return $\overline y_{t}$ as:

\begin{equation}
\overline y_t = \alpha*y_{t-1},
\end{equation}
where $\overline y_t$ represents expected return on time $t$.

\subsection{Unexpected Return}
Next, we would like to compute unexpected return which caused by unpredicted factors, such as temporary market shocks and investor irrationality.
In previous equations, we defined price return consists of expected and unexpected return.
That is:
\begin{equation}
 y_{t} = \overline y_{t} + \theta_{t},~~~t = 1, 2, 3,....
\end{equation}
in which $y_t$ is the real return for TAIEX at time $t$.
$\theta_t$ is the unexpected return, which is the difference between real return $y_t$ and expected return $\overline y_t$.
Meanwhile, we consider the series of unexpected return over the past $N$ days $\Theta_{N,t}$ as:
\begin{equation}
\Theta_{N,t} = \{ \theta_{t-i} | ~i = 1, 2, 3,..., N\}.
\end{equation}
By using historical volatility of unexpected return, we can estimate future price change at time $t+1$ as follows:

\begin{equation}
y_{t+1} = \alpha * y_{t} + \varepsilon,
\end{equation}
where $\varepsilon$ is a white noise $\sim N(0, \sigma^2_{\Theta_{N,t}})$.

Then, we generate the expected price of underlying asset at the expiration date of an option by performing (7) $T$ times which equals to the day remaining before options expiration.
That is:

\begin{equation}
S_T = S_t*\prod_{t=1}^{T} \exp(y_{t}),
\end{equation}
where $S_T$ is the closing price of underlying asset at the expiration date of an option.

\subsection{Monte Carlo Simulations}
Finally, we perform previous processes through MCS for $U$ times and subtract option's strike price from each outcome.
Then, we discount the value with risk-free rate $r$ and calculate the expectation of all traces with (9) and (10), which generate the premium of call options and put options.
\begin{equation}
C = \frac{1}{U}\sum^U_{i=1} e^{-rT} max(S_{T,i}-K, 0),
\end{equation}

\begin{equation}
P = \frac{1}{U}\sum^U_{i=1} e^{-rT} max(K-S_{T,i}, 0),
\end{equation}
where $U$ represents the number of simulation paths in MCS method. 
$K$ is the strike price of the option.
$T$ is the number of days remaining before options expiration.
$S_{T,i}$ stands for $i$-th underlying asset's expected closing price after $T$ days.
Risk-free rate $r$ is 12-months certificate deposit rate announced by Central Bank of Republic of China (Taiwan)
Moreover, we exclude the problem of dividend payment from this paper.

\section{Model Comparison}
In this section, we present the equations of two practiced models, namely Black-Scholes-Merton (BSM) and Binomial Tree (BT), when models are used to price European-style options on a non-dividend-paying index.
Equations are given as follows:

\subsection{Black-Scholes-Merton}

\begin{subequations}
\begin{equation}
C_{BSM} = S_tN(d_1)-Ke^{-rT}N(d_2),
\end{equation}

\begin{equation}
P_{BSM} = Ke^{-rT}N(-d_2)-S_tN(-d_1),
\end{equation}

\begin{equation}
d_1 = \frac{\ln(\frac{S_t}{K}) + (r + \frac{\sigma^2}{2})T}{\sigma\sqrt T},
\end{equation}

\begin{equation}
d_2 = d_1 - \sigma\sqrt T,
\end{equation}

\end{subequations}
where $N(x)$ is the cumulative probability distribution function for a standardized normal distribution.
$S_t$ refers to closing price at time $t$.
$K$ stands for exercise price of the option.
$r$ is the annual risk-free rate using 12-months certificate deposit rate announced by Central Bank of Republic of China (Taiwan).
$T$ is option's annualized time to expiry.
$\sigma$ represents annualized volatility.
$C_{BSM}$ and $P_{BSM}$ denote the price for call options and put options generated from BSM model.

\subsection{Binomial Tree}
From different versions of BT model, we choose the version created by Cox, Ross and Rubinstein, which implies the concept of Risk-Neutral Valuation \cite{1979-Cox-Binomial}.
By using discrete-time approximation to a continuous-time geometric Brownian motion, we can calculate option price $C_{BT}$ and $P_{BT}$ as follows:

\begin{subequations}
	\begin{equation}
	C_{BT} = e^{-rT}*(p*S_tu + (1-p)*S_td),
	\end{equation}
	
	\begin{equation}
	P_{BT} = C_{BT}-S_0+Ke^{-rT},
	\end{equation}
	
	\begin{equation}
	p = \frac{e^{r T} - d}{u - d},
	\end{equation}
	
	\begin{equation}
	u = e^{\sigma \sqrt{T}}, d = \frac{1}{u},
	\end{equation}
	
\end{subequations}

where $p$ represents the probability of the underlying stock to increase following geometric Brownian motion with parameters $r$ and $\sigma$.
$C_{BT}$ and $P_{BT}$ are the price for call options and put options generated from BT model.
The other parameters have the same assumption as BSM model.

Furthermore, annualized volatility used in BSM model and BT model is calculated with N days of historical closing price as follows:
\begin{equation}
\sigma = \sqrt[2]{\frac{\sum^{N}_{i=1}(y_{t-i}-\overline{y})^2}{N-1}}*\sqrt{N},
\end{equation}
where $y_t$ stands for the logarithm of the ratio of closing price at time $t$.
$\overline{y}$ $(= \frac{1}{N}\sum^{N}_{i=1}y_{t-i})$ denotes the average price returns for the past $N$ days. 

\subsection{Performance Indicators}
Finally, we assess the pricing ability among three models with five indicators: Mean error, Standard deviation, NTD Root Mean Squared Error (RMSE) (14), Symmetric Mean Absolute Percentage Error (SMAPE) (15), and Absolute Percentage Error (APE) (16).
Equations are shown as follows:

\begin{equation}
RMSE (NTD) = \sqrt[2]{\sum^Q_{i=1} \frac{(O^{market}-O^{model})^2}{Q}},
\end{equation}

\begin{equation}
SMAPE (\%) = \frac{2}{Q} \sum^Q_{i=1} \frac{|O^{market}-O^{model}|}{O^{market} + O^{model}} * 100,
\end{equation}

\begin{equation}
APE (\%) = \frac{1}{Q * \overline{O^{market}} } \sum^Q_{i=1} {|O^{market}-O^{model}|} * 100,
\end{equation}
where $Q$ stands for the total number of options.
$O^{market}$ is option's market price.
$O^{model}$ denoted the option's closing price calculated by each model.
$\overline{O^{market}}$ represents the average of all options' market price.

\section{Numerical results and discussion}\label{sec:results}

\subsection{Training Period Determination}
In financial market, the fluctuations of volatility can be categorized into short-, mid-, and long-term periods.
These fluctuations were caused by secular trend, irregular fluctuation or periodical adjustment, e.g., seasonal and cycling variation.
Furthermore, previous literature showed that these fluctuations are the reasons that lead to volatility clustering among various financial time series \cite{1997_volatility_clustering}.
In order to adjust volatility clustering, investors usually use a specific period of time, like a month (21 days), a quarter (63 days) or a year (252 days) when calculating annualized volatility.
This process provides investors a reasonable and proper assumption of generating the volatility that reflects market conditions.
Therefore, the annualized volatility applied to BSM model and BT model is calculated from daily historical closing prices based on 252-day rolling horizon statistics and the value of $N$ in MAMC model is set to 252 as well.
\begin{table*}[t]\small
	\centering
	\caption{Performance measurement in three models (2015), U=50000}	
		\label{tb:gaussian_model_test_result}
		\tabcolsep=2.6pt
		\begin{tabular}{lccccccccccccccc}
			\toprule[2pt] &
			\multicolumn{3}{c}{Mean error} &
			\multicolumn{3}{c}{STD} &
			\multicolumn{3}{c}{RMSE (NTD)} &
			\multicolumn{3}{c}{SMAPE (\%)} &
			\multicolumn{3}{c}{APE (\%)} \\
			\cmidrule(r){2-4}
			\cmidrule(l){5-7}
			\cmidrule(l){8-10}
			\cmidrule(l){11-13}
			\cmidrule(l){14-16}
				Option type & MAMC & BSM & BT & MAMC & BSM & BT & MAMC & BSM & BT & MAMC & BSM & BT & MAMC & BSM & BT\\
				\midrule
				All & \bf-5.33 & -5.60 & -12.71 & \bf20.11 & 20.25 & 25.30 & \bf20.79 & 21.00 & 28.30 & 31.35 & \bf31.14 & 47.88 & \bf14.24 & 14.46 & 19.47\\
				\midrule
				Panel A: Type\\
				Call & 0.50 & 0.49 & \bf-0.33 & \bf15.00 & 15.34 & 18.59 & \bf14.99 & 15.33 & 18.57 & \bf27.82 & 27.90 & 31.01 & \bf11.38 & 11.73 & 12.92\\
				Put & \bf-11.16 & -11.69 & -25.10 & 22.73 & \bf22.62 & 25.07 & \bf25.29 & 25.43 & 35.45 & 34.88 & \bf34.03 & 64.26 & \bf16.97 & 17.06 & 25.70\\
				\midrule
				Panel B: Moneyness\\
				ITM & \bf-4.84 & -5.10 & -16.36 & \bf23.13 & 23.05 & 29.99 & 23.59 & \bf23.57 & 34.12 & \bf11.97 & 11.86 & 18.40 & \bf9.62 & 9.66 & 13.92\\
				NTM & \bf-5.98 & 6.27 & -12.46 & \bf20.97 & 21.27 & 24.89 & \bf21.74 & 22.11 & 27.77 & 24.64 & \bf24.33 & 34.48 & \bf19.20 & 19.62 & 24.76\\
				OTM & \bf-5.49 & -5.75 & -9.19 & \bf16.06 & 16.42 & 19.22 & \bf16.94 & 17.37 & 21.28 & 54.12 & \bf53.85 & 84.12 & \bf36.05 & 37.14 & 46.60\\
			\bottomrule[2pt]
		\end{tabular}
\end{table*}

\begin{table*}[t]\small
	\centering
	\caption{Performance measurement in three models (2016), U=50000}	
		\label{tb:gaussian_model_test_result}
		\tabcolsep=2.6pt
		\begin{tabular}{lccccccccccccccc}
			\toprule[2pt] &
			\multicolumn{3}{c}{Mean error} &
			\multicolumn{3}{c}{STD} &
			\multicolumn{3}{c}{RMSE (NTD)} &
			\multicolumn{3}{c}{SMAPE (\%)} &
			\multicolumn{3}{c}{APE (\%)} \\
			\cmidrule(r){2-4}
			\cmidrule(l){5-7}
			\cmidrule(l){8-10}
			\cmidrule(l){11-13}
			\cmidrule(l){14-16}
				Option type & MAMC & BSM & BT & MAMC & BSM & BT & MAMC & BSM & BT & MAMC & BSM & BT & MAMC & BSM & BT\\
				\midrule
				All & 5.50 & 3.85 & \bf-3.56 & 21.69 & \bf21.52 & 27.08 & 22.36 & \bf21.85 & 27.29 & 31.44 & \bf30.92 & 44.19 & 16.76 & \bf16.32 & 20.43\\
				\midrule
				Panel A: Type\\
				Call & 14.53 & \bf12.91 & 13.34 & 17.77 & \bf16.99 & 17.30 & 22.94 & \bf21.32 & 21.83 & 42.18 & \bf40.29 & 41.68 & 20.77 & \bf19.01 & 19.50\\
				Put & \bf-3.53 & -5.21 & -20.46 & \bf21.51 & 21.77 & 24.41 & \bf21.77 & 22.36 & 31.83 & \bf20.71 & 21.56 & 46.69 & \bf13.46 & 14.09 & 21.21\\
				\midrule
				Panel B: Moneyness\\
				ITM & 5.81 & \bf4.23 & -6.46 & 26.30 & \bf25.90 & 32.56 & 26.90 & \bf26.20 & 33.15 & 13.43 & \bf13.19 & 16.17 & 11.66 & \bf11.36 & 14.25\\
				NTM & 7.13 & 5.12 & \bf-1.76 & \bf20.76 & 20.80 & 26.53 & 21.89 & \bf21.35 & 26.50 & 25.44 & \bf24.86 & 31.34 & 21.85 & \bf21.07 & 25.73\\
				OTM & 4.38 & 2.83 & \bf-1.57 & \bf16.41 & 16.48 & 20.28 & \bf16.96 & 17.00 & 20.31 & 52.47 & \bf51.71 & 78.64 & 42.53 & \bf41.59 & 52.81\\
			\bottomrule[2pt]
		\end{tabular}
\end{table*}

\subsection{Simulation Results}
After determining the training period, we use MAMC model to price options which was issued in 2015 and 2016.
These options were also evaluated via practiced models.
The results of option pricing in 2015 and 2016 were shown separately in TABLE 3 and TABLE 4.
Moreover, we classified options based on their (A) Type (Call/Put), and (B) Moneyness to investigate further implications under different scenarios.

First, the outcomes in 2015 are different from the results in 2016.
In 2015, MAMC model has the best performance; however, BSM model is a better solution in 2016.
Despite the fact that no model has absolute advantage in short-term options pricing, this result shows that MAMC model has the ability to price short-term options as well as practiced models.
In addition, BT model has the poorest outcomes among three models.

Second, we investigate the results under different scenarios in Panel A and Panel B.
In Panel A, we report the pricing performance for each model upon different option types (Call Options/Put Options).
While determining the premium of put options, the outcomes show that MAMC model outperforms other models in both 2015 and 2016.
Despite the significant outcomes on put options pricing, MAMC model shows a modest ability to evaluate call options.
In 2016, BSM model has better performance on call options pricing.
These results point out the advantage of our approach while pricing put options.

Then, we separate options according to its moneyness, which represents a situation whether an option is generating profit.
The results show that MAMC model outperforms other models on most parts (12 out of 15) in 2015.
On the other hand, BSM model has better performance on the mass of options (10 out of 15) in 2016.
After taking a deeper look into three segments in Panel B, we find out that the results of both STD and RMSE show a decreasing trend from ITM options to OTM options; however, SMAPE and APE, indicators of error measurement in percentage scale, have an increasing movement.
While these phenomenon occur in all models, we believe that the outcomes are caused by different types of moneyness.
After investigating this matter, we observe the fact that market prices of OTM options are usually lower than market prices of ITM and NTM options.
In some extreme cases, price of OTM options were lower than one.
These phenomenon will shrunk down the denominator in the equations of SMAPE and APE and lead to poor performance on evaluating OTM options.
Hence, RMSE is a better indicator for comparing the performance of OTM options pricing.

In order to evaluate overall performance of three models, we decide to check which model has the advantage over other models when pricing each type of options (All, Panel A and Panel B).
To sum up, MAMC model performs better (8 out of 12) than BSM model (4 out of 12) on pricing weekly options within 2015 and 2016.
\section{Conclusions}
\label{sec:conclude}
In this paper, we provide a compound model for option pricing via AR process and MCS method.
With fine results yielded by MAMC model, we prove the feasibility of applying this methodology to price short-term options.
However, some disadvantages exist in MAMC model when pricing call options.
These results are the direction of our future researches.
First, we will probe into order selection on AR process in order to improve MAMC model's performance.
Second, MAMC model shows worse results when evaluating call options in 2016 than in 2015.
The reasons of worse outcomes might need further investigations.
Finally, MCS method serves as the basis of advanced Machine Learning models (e.g., ANN and Deep Learning).
Using these models in the field of option pricing could be another interesting direction.
\vfill\pagebreak
\bibliographystyle{unsrt}
\bibliography{./References_ar_project}
\end{document}